\documentclass{article}
\usepackage{amsmath}
\usepackage{hyperref}

\usepackage{algpseudocode}

\AtBeginDocument{%
  }

\begin{document}

\title{Multiplying matrices using $n$ arithmetic operations}

\author{Hugo Daniel Macedo}
\maketitle

\begin{abstract} 
It is widely known that the lower bound for the algorithmic
complexity of square matrix multiplication resorts to at
least $n^2$ arithmetic operations. The justification builds
upon the following reasoning: given that there are $2 n^2$
numbers in the input matrices, any algorithm necessarily
must operate on each at least once. In this paper, we show
that this is not necessarily the case for certain instances
of the problem, for instance matrices with natural number entries. We present an
algorithm performing a single multiplication and $(n-1)$
sums, therefore using $n$ arithmetic operations. The
ingenuity of the approach relies on encoding the original $2
n^2$ elements as two numbers of much greater magnitude.
Thus, though processing each of the inputs at least once, it
relies on a lower count of arithmetic operations. In
the computational model used to analyse this problem, such
encoding operation is not
available, thus it is not clear this work affects the
currently accepted complexity results for matrix multiplication, but
the new algorithm complexity (when taking into account the
encodings) is
$3 n^2 + 2 n - 1$ operations. In addition, given the
exponential increase in multiplication operands magnitude,
its practical usage is constrained to certain instances of
the problem. Nonetheless, this work presents a novel
mathematically inspired algorithm while pointing towards an
alternative research path, which opens the possibility of novel
algorithms and a taxonomy of matrix multiplications and
associated complexities. 
\end{abstract}

\section{Introduction}

No matter where one finds the topic of the algorithmic
complexity of square matrix multiplication, matrices of size
$n \times n$, one reads the operation has a quadratic lower
bound. For instance, in the recent work of
\cite{Williams&23} one finds the current bound and an
account on the many decades of study of this problem, where
the problem is classified as $\mathcal O(n^{\omega+\epsilon})$,
where $\epsilon > 0$ and $2 \leq \omega < 3$.
In \cite{Harvey&18}, the complexity of integer matrix
multiplication is dominated by a term  of $n^2 M(p)$, where
$M(p)$ denotes the bit complexity of multiplying $p-$bit
integers.
In the work of
\cite{Raz&03}, a similar and broader account is given, with
an additional emphasis on the ``standard computational model for
matrix product'': arithmetic circuits over some field. 
Given that the only allowed operations in such circuits are
arithmetic operations, it is easy to prove that any
algorithm  must at least invoke $2 n^2$ arithmetic
operations.
In this paper, we show that this is not necessarily the
case, by ``thinking outside the box'' such computational
model enforces. At least for a large number of instances of
the problem.


The new approach relies on encoding and
decoding operation of the original matrix inputs and
outputs into numbers as a standard computer would load and
shift entries from memory into the registers of its
arithmetic units. By choosing the
right encoding, it is possible to devise an algorithm using
$n$ arithmetic operations in the case of positive entries
from certain fields. The encodings are naturally
arising from the study of matrix multiplications in the
framework of the universal law for tensor products, for
instance as detailed in 
\cite{Macedo&16}, but such mathematical framework requires
a technical detour irrelevant for the understanding of the
inner workings of the new algorithm. 

For the reader plainly curious about the result, for
instance, a freshman linear algebra student,
we illustrate the algorithm in Section \ref{sec:il}. Given
the need to take into account the entries magnitudes and
perform encodings tailored to each dimension, we present a
generic template to devise matrix multiplication algorithms
in Section \ref{sec:af}. Each instance of such algorithm can
be analysed in terms of complexity in Section \ref{sec:ca}
and a brief discussion of the implications the new algorithm
(and future extensions) brings to the
complexity results around this problem are given in Section
\ref{sec:imp}.


\section{Illustration of the algorithm}
\label{sec:il}

To easily understand the approach, and because this article
intends to also serve a broad audience interested in
algorithms, let
us focus on a simple ($n=2$) instance of the problem, where
the inputs and output feature bounded magnitude entries
(all entries can be represented with 2 digits). Let us
assume the following product of matrices is to be computed:
\[
\begin{bmatrix}
    1 & 2 \\
    3 & 4 
\end{bmatrix}\cdot 
\begin{bmatrix}
    5 & 6 \\
    7 & 8 
\end{bmatrix}= 
\begin{bmatrix}     19 & 22\\     43 & 50 \end{bmatrix}
\]
The first step consists in transforming the multiplication
of the input matrices into a multiplication of two
single numbers encoding the original matrices:
\[ \begin{bmatrix}     1 & 2 \\     3 & 4  \end{bmatrix}
\cdot \begin{bmatrix}     5 & 6 \\     7 & 8  \end{bmatrix} 
\]
The encoding vectorizes the left-hand matrix in row-major
and the right-hand one in column-major order\footnote{The
template algorithm described in the paper requires different vectorizations due to
the need to cope with general dimensions and digit
sizes\ldots Nonetheless for these two $2\times 2$ matrices, it becomes more
concise and understandable to use this approach.}. 
\[
0102010203040304 \cdot 0507060805070608
\]
Notice that up to this point no arithmetic operations were
used and how the vectorization builds a vector as a number
representation by introducing zeros instead of the usual
vector boxing/spacing. At this point, perform the next
step and compute the
multiplication of the two encodings, resulting in:
\[
0514061615281832
\]
The following step is to perform the  $n-1$ additions, in this case
a single addition for the matrix dimension is two:
\[     0514061615281832 + 051406161528183200 \]
At this point, the reader should notice how the right-hand side
operand was shifted two digits to the left, as it is
important to add to the count of operations. Yet another
non-arithmetic operation available in any computational
device.
The result of the two arithmetic operations result in the
following number:
\[
     5{\bf 19}20{\bf 22}31{\bf 43}46{\bf 50}32
\]
Which can be decoded into matrix form, obtaining the
expected result:
\[
\begin{bmatrix}
    19 & 22\\
    43 & 50
\end{bmatrix}
\]
Thus performing the matrix multiplication using two
arithmetic operations only. If the interest is the total of
operations used, one may argue 12 operations for
encoding/decoding, one multiplication, one sum, and one
shift. Thus consistent with the $3 n^2 + 2 n -1$ formula.

\section{A family of matrix multiplication algorithms}
\label{sec:af}
For  any two square matrices $A$, and $B$ containing natural $n
\times n$ entries that lead to at maximum
$p$-sized digits in the resulting matrix $C$, it is possible
to devise an algorithm to perform such 
matrix multiplication using $n$ arithmetic operations by
instantiating the following template:

\begin{algorithmic} 
    \Function{MMM$_{p,n}$}{$A$,$B$} 
    \State $a \gets encode^{lhs}_{p,n}(A)$ 
    \State $b \gets encode^{rhs}_{p}(B)$ 
    \State $x = a \cdot b$ 
    \For{$k = n-1$ to $1$}  
    \State $x = x + shiftl_{p,n,k}(x)$ 
    \EndFor 
    \State $C \gets decode_{p,n}(x)$ 
    \State \Return $C$ 
    \EndFunction 
\end{algorithmic}

The template function depends on expanding the following
template auxiliary functions:

\begin{itemize}
    \item $shiftl_{p,n,k}(x)$ appends the number $x$ with
        $10^{(k * (p*(n^2+1)))}$ zeros
    \item $encode^{lhs}_{p,n}(M)$ pads each of the elements of the row-major vectorization
        of matrix $M$ with zeros to obtain a representation
        occupying $n^2\times p$-digits
    \item $encode^{rhs}_{p}(N)$ pads each of the elements of the
        column-major vectorization of matrix $N$ with zeros to obtain
        a representation occupying
        $p$-digits
    \item $decode_{p,n}(x)$ obtains the $C_{ij}$ element from the
    number $x$, by transforming it into a word $w$ of 
    size multiple of $p$  and containing all the digits of
    $x$, and selecting the digits between
$w_{((i\cdot n \cdot p-p)\cdot n^2 + j \cdot n \cdot p-p)}$
and $w_{((i \cdot n \cdot p-p)\cdot n^2 + j\cdot n\cdot
p)}$ 
\end{itemize}

At this point, after a careful analysis, the reader may find
itself disappointed by the generic heading of the
article, as the algorithm shown works only when applied to
natural number entries. Nonetheless, this algorithm is
enough to claim a new branch to this problem could be
developed. It seems possible to extend the algorithm
to cover truncated positive real entries. With careful
tracking of signs and the inclusion of a subtraction after
each sum\footnote{Or using an implicit sign number
representation.}, it should also possible to further cover
the negatives in that subset. Whether the previous claims
hold or it is possible to cover more fields with
extensions is outside the scope of this article, but the
questions draft a path towards a taxonomy of matrix multiplication
problem instances and their complexity. Though the
algorithm brings measurable performance improvements for
small matrices with tiny magnitude entries, the main focus
is the theoretical construction based on encodings and its
theoretical implications.

\section{Complexity analysis}
\label{sec:ca}

For each problem size $MMM_{p,n}(A,B)$ the algorithm
resulting from instantiating the template performs one
multiplication and $n-1$ addition arithmetic operations. The
$encode$ and $decode$ functions require $3 n^2$ operations,
we assume each entry encoding can be performed by a single
operation. For instance, a standard load/move operation in
computational devices. Analogous reasoning is applied for the $n-1$ instances
of the $shiftl$ operations. The total so far amounts to $3 n^2 + 2 n
-1$ operations and classifies the algorithm in $\Omega(n^2)$
operations in an ad hoc extension of the standard algebraic
computational model. As the complexity of load/move
operations are typically discarded in complexity theory, we
can pose the algorithm in $\Omega(n)$ (arithmetic)
operations. The extensions to cover the whole field of
reals add at least $(n-1)$ subtractions, and the derivation
of the exact formula is tied to each particular extensions,
thus left as future work. 


\section{Concluding remarks}
\label{sec:imp}

The previous works posing lower bound works as $\omega \geq
2$ remain unchanged. There is no mistake in the previous
conclusions, given that all  arguments assume as a premise
the algebraic computational model. Moreover, the presented
algorithm lacks the original works generality, which covers any
field.  Yet, given the possibility of using only $n$
arithmetic operations to multiply matrices, it seems
unreasonable to accept such premise as the departure point
for the deeper study of matrix multiplication. 

In case it is acceptable this algorithm provides an
effective linear run time, $\Omega(n)$ arithmetic
operations, it would be possible to extend the standard
algebraic model of computation to include encodings.
Interestingly, the algebraic computational model is taken as
too powerful (given the ability to multiply any number of
arbitrary magnitude), but as we observe in this study, it is
also  too restrictive,  given the inability to model the
encoding/decoding and shift operations without resorting to
the arithmetic operation gates. With the inclusion
of number encodings, it may be possible to find an adequate
computational model for algorithmic innovations in other
areas.

Though the results are enough to make an advance in the
search 
for matrix multiplication algorithms, the current algorithm
is restricted to certain subsets of the generic algorithm,
so further work is expected to assert the exact generic
complexity and how to relate it with the current results
established in terms of non-natural fields.

The same encodings allow us to multiply two polynomials
using a single multiplication plus $3 n$ encoding/decoding
operations. Nonetheless, the research in such domain is left
for future work, because polynomials are themselves used to 
implement the fastest multiplication \cite{Harvey&21} known. Given such
recursive nature, one can quickly achieve a number of 
multiplication algorithms that are interesting from a
theoretical point of view, but the ingenuity of encoding
numbers as polynomials and using point evaluation to improve
running times seems far more efficient than recursively
calling a multiplication algorithm with two numbers of
exponentially larger magnitude.

\newpage 

\bibliographystyle{ACM-Reference-Format}
\bibliography{mmm}

\end{document}